# The Recognizability of Authenticity

**Madeleine Henderson (madeleine.henderson@gmail.com) and Liane Gabora (liane.gabora@ubc.ca)**
Department of Psychology, University of British Columbia
Okanagan Campus, 3333 University Way, Kelowna BC, V1V 1V7, CANADA

**Abstract**

The goals of this research were to (1) determine if there is agreement both amongst viewers, and between viewers and the performer, about the extent to which performances are authentic, and (2) ascertain whether or not performers and/or viewers can distinguish between authenticity and skill. An authentic performance is one that is natural or genuine, while an inauthentic performance feels faked, forced, or imitative. Study participants were asked to rate the authenticity and skill level of a series of videotaped performances by dancers and stand-up comedians. Performers also rated their own performances. Authenticity ratings amongst viewers were significantly positively correlated. Ratings between viewers and performers were not significant but all positive. A higher correlation between ratings of both authenticity and skill of performances for viewers than for performers suggests that viewers make less of a distinction between authenticity and skill than performers. The relationship between authenticity and creativity is discussed.

## Introduction

With increasing frequency there are calls for research aimed at a synthetic account of how the components of a cognitive system function in synchrony to generate behavior in everyday situations. We propose that the construct of authenticity has an important role to play in such an account. Authenticity refers to the ability to be genuine, to accurately reflect who one really is, and be true to the situation one is in. Writers speak of discovering one's own authentic voice. In theatre research the term 'authentic' is used in discussion of the extent to which a performer gives a performance something personal that goes beyond the script (Lavy, 2005). In the dance community the term 'authentic movement' refers to the strengthening of identity through uninhibited movement of they body in a social context (Goldhahn, 2009). In an area at the intersection of anthropology and tourism research, the term 'authenticity' is used to refer to the extent to which current creative works in a given genre, such as Native American or First Nations art or dance, employ the same tools, techniques, styles, and so forth, as were traditionally used (Daniel, 1996; Maruyama *et al*., 2008). Thus an authentic performance is one that seems natural, or true to an underlying essence, while an inauthentic performance feels faked, forced, or imitative.

Authenticity is important for many reasons. It feels highly gratifying to both the performer and the observer. It is relevant to many domains of life, including the generation of artistic works and performance (*e.g*., art, acting, music, and dance), non-artistic performances (*e.g*., teaching and newscasts), and everyday social interactions with friends and family. However, despite that performers, viewers, and the general public regularly voice opinions about authenticity, and despite that in the scholarly community authenticity is assumed to be a genuine construct about which viewers and performers are in agreement (*e.g*., Goldhahn, 2009; Kogan, 2002; Lavy, 2005; McClary, 2007; Nemiro, 1997; Sawyer, 1992; Warja, 1994), we were unable to locate any empirical research that supports this assumption. Indeed we found no empirical evidence for consensus as to which performances are authentic and which are not, either amongst members of an audience, or between a performer and an audience.

### Authenticity and Skill

Audiences without artistic expertise emphasize skill over originality in assessments of visual art, while the reverse is true for audiences with expertise (Hekkert & van Wieringen, 1990a, 1990b, 1996). This suggests that originality—which might be related to authenticity—can be confused with skill. However, there is evidence that skill and authenticity are distinct constructs (Kogan, 2002). While being skilled in a domain may facilitate authentic performance, it does not guarantee it, nor is it a necessary prerequisite. For example, a dancer may have perfected her craft, and be technically skilled, permitting a wide range of means for self-expression, but not immerse herself in the work, or simply imitate the instructor, yielding a performance void of authentic style. Conversely, a performer lacking in technical skill may exude personality or detectable "creative release", yielding a performance that comes across as authentic. In short it remains an open question whether viewers confuse a skilled performance with an authentic one.

### Goals of Current Study

Although it would be difficult to pinpoint the potentially myriad factors that contribute to authenticity or a lack of it, it is possible to make headway toward determining whether authenticity is a genuine construct by assessing the extent to which viewers of a performance, and performers themselves, agree in their assessments of authenticity. Thus a first goal of this study was to determine if there is a correlation amongst viewers' assessments of the authenticity of a given performance. A second, related goal was to determine whether there is a correlation between viewers' assessment of the authenticity of a performance and the performer's self-assessment of the authenticity of that performance. We hypothesized that an audience can detect an authentic or inauthentic performance, and that performances that feel authentic to a performer come across as authentic to an audience, and *vice versa*.

A third goal was to determine whether authenticity and skill are distinct constructs in the eyes of the performer and/or viewers. Since it is possible to be skilled but perform in an inauthentic manner, or to perform authentically but not be skilled, we hypothesized that both performers and viewers could distinguish between the two constructs.

A final goal was to determine what factors facilitate authentic performance. Previous research on this is inconclusive (e.g., McClary, 2007; Nemiro, 1997; Rhodes, 1999; Sawyer, 1992; Warja, 1994). By asking performers open-ended questions about authenticity we hoped to shed light on this seemingly elusive phenomenon that would pave the way for further studies of the relationship between authenticity and the therapeutic value of creative endeavors.

## The Study

### Participants

Three trained dance performers were recruited from a local dance studio. Dancer A was 25, Dancer B was 29, and Dancer C was 23. Each dancer had between 10 to 12 years of dance experience, and took part in dance at least once per week. Each trained dancer was paid $30 for their participation in the study. They met with the experimenter for a video recording session of three hours duration. A four year old child with no formal dance training was also recruited as a dance performer.

Three comedians were also recruited for the study. The first was a 36-year old experienced stand-up comedian with eight years of stand-up comedy experience. He was located from a local directory. The second was a 24-year old amateur stand-up comedian who had just started doing stand-up comedy one month prior to the study. She was recruited through a psychology of humour class at The University of British Columbia. The third was a 23-year-old 'social comedian' known to the experimenter. He had no stand-up comedy credentials, but had years of experience being the center of attention for his humour in social situations. None of the comedians were compensated for their participation.

158 University of British Columbia undergraduates were viewers of the performances. 45 were recruited through the SONA system, which enables participation in university research in exchange for credit in a psychology class. 50 students were recruited through psychology of creativity and psychology of humor classes. They were not given incentives to participate. The remaining participants were recruited through online university class message boards, and were also not given incentives to participate. The only exclusion criterion was severe visual impairment, such as blindness. Females accounted for approximately 58.2% of the sample (n=92) and males accounted for the remaining 41.8% (n=66). Most (83.1%, n=128) were between the ages of 17-25, and in a Bachelor of Arts (64%, n=96) program.

### Procedure

The experienced dancers were filmed practicing original choreographed dance routines in their dance studio. They were told that the study was about the psychology of movement. They met at the dance studio one hour prior to filming to learn two different modern dance routines. Both routines were choreographed to music and lasted one to two minutes in duration. For one, the music was a quick, high-energy piece, while for the other it was slow and sombre.

After the hour-long practice, each dancer individually performed the fast dance five times. After all dancers had finished, they individually performed the slow dance five times in the same order as the first. Each dance performance was videotaped using a high-definition video camera.

Once the first dancer had completed all her dances, she was directed to a laptop where footage of her routines was uploaded. She was debriefed about the specific reasons for conducting the study, and given a definition of authentic performance. She was then asked to watch her own ten performances in the order in which they were performed, and given a questionnaire with the following items based on each performance:

> Please rate how authentic you felt this performance was based on how you felt you were coming across or how you felt inside during the performance" (Not authentic at all / Somewhat authentic / Neutral or Don't know / Quite authentic / Very authentic)
>
> How would you rate your performance in regards to technical skill?" (Very poor / Poor / Okay / Good / Very good)

This was repeated for the other dancers. All dancers were also asked to fill out an open-ended portion of the questionnaire, which asked the following questions concerning factors that facilitate or hinder authenticity:

(1) Do you feel as though authenticity and technical skill are the same thing or different concepts? Please explain.

(2) Are there particular situations or environments in which you are able to produce your most authentic performance? If so, please tell us about it.

(3) Do you believe that people become more able to find their authentic style with experience?

(4) Was there a known time in your career where you felt that you had made a transition to being more able to express yourself? If yes, describe that transition.

The child dancer was filmed using a high-definition video camera in her home. Filming began when she spontaneously began dancing to upbeat dance music. The camera was not hidden from view and the child was aware she was being filmed. The footage was divided into two video clips of two minutes each. Due to her age, she was not asked to assess how authentic or skilled her performances were, nor to respond to the open-ended questions.

The experienced stand-up comedian was asked to submit between five and ten previously taped performances that he had acquired over his career. We requested that each video clip be under two minutes duration, and that together they portray a range of authenticity. He was also asked to rate the authenticity and skill of each performance using the same five-point Likert scale administered to the dancers. He submitted six videotaped live performances and his ratings for each.

Video footage was collected from the amateur stand-up comedian without an audience (except for the experimenter). She was asked a series of questions that would potentially promote humourous responses, such as “what was your most embarrassing moment?” or “what is the strangest thing you have seen on campus?” She was also asked to run through segments of her stand-up routine which were narrative in nature, tell funny jokes (not necessarily her own), or make up funny stories and deliver them as though they were real. After approximately 30 minutes of recording, she was asked to look through the footage on a laptop and rate the authenticity and skill of each joke/story segment on the five-point Likert scale. Video footage of the social comedian was collected in the same manner as with the amateur stand-up comedian.

The video clips were loaded onto an online questionnaire using www.surveymonkey.com with the exception of those from Dancer C. Her performances were omitted due to extreme homogeneity in her responses to the Likert items. (Since her performances did not exhibit variation in self-rated authenticity, they were not useful for this study.) Her responses to the qualitative questions were retained.

Viewers were given the following definition of authenticity:

> Authenticity in the performing arts commonly refers to the ability of a performer to perform in such a way that they are able to remain true to who they really are or to the character they are trying to play. Conversely, a performer who is not performing authentically is merely giving a performance that seems artificial or imitated.

Viewers were asked whether they felt they understood the construct of authenticity, and if they did not, further discussion ensued until it was clear to them what authenticity refers to. After each video clip, viewers were required to rate it on the same five-point Likert scales that the performers used. In order to minimize potential order effects, the ordering of the performances was randomly altered every time ten students had completed the survey. Video clips belonging to the same performer were kept together, but the order of the performers and the order of the clips belonging to each performer were randomized.

The students who were recruited from the psychology of creativity and psychology of humor classes were shown the video clips on a projector screen, and they received a paper version of the questionnaire. They were given 30 seconds to rate each performance before the next one commenced. The type of psychology class and the week in which the study was conducted determined the types of performances that were shown. For example, the psychology of creativity class was approached earlier in the study, and was shown the clips of the dancers and the experienced stand-up comedian because these performances were the only ones available at that time. The psychology of humour class saw only the comedians’ performances because dance performances were not relevant to the class content.

## Analysis and Results

The means and standard deviations for the authenticity ratings of the performances are given in Table 1. The highest authenticity ratings were for the dancing child ($M$ = 4.52) and the social comedian ($M$ = 4.05).

Table 1: Mean authenticity ratings by viewers and performers for all performances.

| Performer | Viewer ratings *M (SD)* | Performer Self-ratings |
|---|---|---|
| **Experienced stand-up comedian** | 3.68 (1.07) | 3.33 |
| **Amateur stand-up comedian** | 3.19 (1.31) | 2.50 |
| **Social comedian** | 4.05 (1.08) | 4.33 |
| **Dancer A** | 3.18 (1.12) | 3.80 |
| **Dancer B** | 3.27 (1.18) | 3.30 |
| **Dancing child** | 4.25 (1.01) | N/A |

### Recognizability of Authenticity

**Between-Viewer Ratings** To determine whether the viewers agreed as to which performances seemed authentic, the intraclass correlation coefficient ($R_i$) was calculated. The $R_i$ statistic is more appropriate for this study than the widely-used Pearson product moment correlation because the latter ignores the extent to which independent raters agree on any single rating (Cicchetti, 1991). The $R_i$ coefficients for the extent of agreement amongst viewers about the authenticity of the performances of each performer are presented in Table 2. All values are statistically significant at the .05 level with the exception of those for Dancer A, and they are all statistically significant at the .01 level with the exception of those for Dancer A and the dancing child.

Table 2: Agreement of authenticity amongst viewers ($R_i$), and between viewers and performer ($r$).

| Performer | $R_i$ | $r$ |
|---|---|---|
| **Experienced stand-up comedian** | .965** | .712 |
| **Amateur stand-up comedian** | .890** | .120 |
| **Social comedian’** | .858** | .609 |
| **Dancer A** | .340 | .061 |
| **Dancer B** | .879** | .520 |
| **Dancing child** | .822* | N/A |

*$p$<.05; **$p$<.01

**Agreement Between Viewers and Performers.** To determine if there was agreement between viewer and performer ratings of authenticity, we merged the multiple viewer ratings to obtain the average composite rating for each performer. A Pearson product moment correlation was conducted to see if the composite rating is in agreement with the performer's ratings of authenticity. These values are also presented in Table 2.

The highest agreement was between the viewers and the experienced stand-up comedian, followed by the social comedian, Dancer B, the amateur stand-up comedian, and Dancer A. There was considerable variation amongst the performers with respect to the degree to which their assessments of the authenticity of their performances were correlated with the viewers' assessments. While none of the correlations were statistically significant, all were positive. Moreover, significance was based on a small number of performances for each performer. The lack of power from the small *n*'s indicates that the significance tests were highly prone to type II errors (failure to find a significant difference when one exists). In such situations it may be prudent to focus on the magnitude of the observed effect or relationship instead of the significance tests (Gliner, Leech, & Morgan, 2002; Serline & Lapsey, 1993; Wilkinson & the APA Task Force on Statistical Inference, 1999).

**Qualitative Results.** To better understand what factors facilitate the expression of authentic creative style we conducted a content analysis of the open-ended questions. There were recurring responses as well as individual differences. Responses to the question, "Is the development of an authentic voice related to experience?", suggest that experience facilitates the development of authentic style, but that this happens differently for different performers. Compare the responses of two dancers:

> Experience is what helps one explore his or herself to discover what authenticity means for them.

> I have found that by taking a number of different dance styles with a number of different instructors that I have developed (and continue to develop) my own personal style. The more experience that I've gained the more comfortable I've become with myself and my movement and the more ideas that I can "pull out of my hat".

The performers put forward several factors that interfere with the authenticity of their performances: excessive focus on technical perfection, performing in front of large audiences, or audiences that include friends or acquaintances, performing while injured or tired, performing content that is unfamiliar or that does not "lean towards [one's] natural expression", and working with a choreographer that has a different style. The performers also put forward many factors that enhance with the authenticity of their performances. The most commonly cited factor was feeling safe from judgment. Other factors were being in a performing mood, feeling inspired, and teaching choreography. Interestingly, while some performers claimed that having an audience increases the authenticity of their performance, others claimed that it has the opposite effect.

## Distinguishing Authenticity from Skill

**Quantitative Results**. There was a modest but significant Pearson correlation between mean ratings of authenticity and mean ratings of skill as assessed by viewers. The Pearson correlation for the performers' mean ratings of authenticity and skill was lower but significant. These results are presented in Table 3. Thus although authenticity and skill appear to be related for both performers and viewers, performers made a stronger distinction between them than viewers.

Table 3: Pearson correlation between mean ratings of authenticity and mean ratings of skill (*r*).

| | *r* |
|---|---|
| **Viewers** | .641 |
| **Performers** | .547 |

$^*p<.001$

**Qualitative Results**. The qualitative data indicates that the performers unanimously view authenticity and skill as distinct concepts. For example:

> Technical skill – is where you learn how to move and hold yourself properly for the desired discipline. Authenticity – is the feeling and expression that you can add to your technical skill to create the "entire picture".

> Anybody can master technical skills with enough practice but if you don't have charisma as an artist – or better yet as a stand-up comedian, people won't think you're very funny.

Responses suggested that skill can facilitate authenticity:

> Technical skill opened the door of possibilities for me to further express my emotions.

> Once I know how to do a proper "plie" and the barre, it is much easier for me to add some expression or feeling because I'm not thinking nearly as much about how the plie should be done and can focus on making it look "pretty."

However, one performer's answers suggested that acquiring skill may *interfere* with authenticity:

> Sometimes a lot of technical training can make it difficult for the dancer to separate their own authentic style from the teachers. It all comes down to how they have been trained, if their teacher demands uniformity and discourages personal exploration it will be harder. If they have a good teacher who knows how to pull out creativity and massage it, then the experience will benefit their discovery of an authentic style.

The performers claimed that skill can facilitate authentic performance by freeing them from concern with technical details so they could be more fully immersed in the creative process. A preoccupation with skill, however, can prevent a performer from reaching a deeper connection with the task. These qualitative responses, in conjunction with the quantitative results, support the hypothesis that authenticity and skill are related, yet distinct concepts.

## Discussion

The results of this study shed light on the seemingly elusive construct of authenticity. The agreement amongst viewers as to which performances were authentic, a result obtained across a variety of performance types and situations, suggests that authenticity is indeed a real concept as opposed to existing in the eye of the beholder.

The variability in the correlations between authenticity ratings for viewers and performers indicate that when a performance feels authentic to a performer it may or may not come across that way to others. This was addressed by one of the dancers, who noted:

> Some people have very 'quiet' personalities so when they are authentically displaying anger they might be so quiet about it [that] an audience would not see it. Those dancers might be rated 'less authentic' because they are less obvious.

This comment suggests to us that the reason for the low agreements between the amateur stand-up comedian and Dancer A and the viewers is that outward manifestations of their personalities may be subtle for the viewers to detect them. Analyzing how the personality of an artist interacts with the recognizability of authenticity in performance is an interesting direction for future research.

Although the variety of performance types and settings contributed to the generalizability and ecological validity of the findings, caution must taken in drawing conclusions that involve comparisons *across* performers or performance settings, because differences such as 'in a studio' *versus* 'at home' could be potential confounds. With this warning, we offer some speculative discussion of between-performer differences. There are several possible explanations for the high agreement amongst viewer authenticity ratings of the experienced comedian. First, over time he may have solidified a strong authentic voice that is readily detectable when present, making an inauthentic performance stand out in contrast. Second, his performances were the only ones that were filmed before he knew he would be rated. Some research indicates that the pressure of knowing one is going to be evaluated can inhibit creative expression (Nemiro, 1997; Rhodes, 1999), so it is possible that the rest of the performers who knew they were going to be evaluated gave performances that were more uniform with respect to authenticity, giving viewers less opportunity to detect differences amongst performances. It would be interesting to investigate whether expertise can entail becoming skilled at *faking authenticity*, *i.e.,* whether there exist performers for whom expertise is inversely correlated with agreement between performer and viewer authenticity ratings.

The dancing child's high authenticity ratings may reflect in part the stereotype that children are authentic in whatever they do. However, the fact that the social comedian's performance was also rated as highly authentic suggests that these high ratings reflect instead the spontaneity of their performances. While the other performers' performances (though to a lesser extent the amateur comedian) were choreographed or scripted, those of the child and social comedian were not. This interpretation is consistent with findings that freedom facilitates authenticity (McClary, 2007; Nemiro, 1997; Sawyer, 1992; Rhodes, 1990, Warja, 1994). This explanation is further reinforced by the fact that the experienced stand-up comedian's highest rated performance for authenticity was the only one in which he was forced to improvise (due to verbal feedback from the audience). This points to a weakness of the study. Since most performers knew they would be judged, they may have been less able to release inhibitions and be authentic. Another weakness is that because dancers were confined to rehearsed, choreographed routines, differences between performances of the same dance may have been too subtle for viewers to detect, thus limiting the range of authenticity scores. Future studies will focus on spontaneously improvised performances, which allow authenticity to be expressed through content as well as delivery.

Another direction for future research is to investigate whether there is a difference in the capacity to detect authenticity in live versus videotaped performances. Previous research indicating that there is a constant interaction between a performer and a live audience (Arnold, 1991; Bindeman, 1998; Nemiro, 1997) suggests that viewers may be better able to detect cues or indications of authenticity from a live performance than from a performance on a television or computer screen.

This study of authenticity arose in the context of an interest in what factors affect how the various components of a cognitive system come together to produce overt thought and behavior. It seemed reasonable that an authentic response is one that genuinely reflects the state of one's associative network, including not just one's internal model of the world (including self-understanding) but one's way of seeing and being. We refer to this dynamical structure as a *worldview* (Gabora, 1999, 2000, 2001). We speculated that: (1) authenticity entails being entirely present and thereby available to detect unresolved questions or issues, and open to change, (2) a lack of authenticity may indicate that elements of one's worldview are repressed or misrepresented because this interferes with the capacity to detect unresolved questions or issues, and be open to change, (3) authentic performance facilitates the process by which one's worldview self-organized into more stable state, while unauthentic performance does not. However, before we could address these issues it was necessary to address the more fundamental question of whether authenticity is a real construct.

The findings reported here, and in particular the key finding that authenticity is recognizable, opens up many questions and perspectives. It led us to speculate that perhaps one is being creative even when not engaged in an overtly creative activity if one responds to a new situation or emotion in a way that authentically reflects how it affects ones' worldview. This is consistent with the honing theory of creativity, according to which creative behavior arises because one's worldview tends to self-organize in response to perturbation to achieve a more stable state, regain equilibrium, or resolve dissonance (Gabora, 2005; Gabora, Ranjan & O'Connor, 2012). The creative product or performance is viewed as an external reflection of this internal transformation. This conceptualization of creativity is consistent with the anecdotal evidence obtained in the qualitative portion of this study that authentic performance can be therapeutic. It is also consistent with the notion that personal performance style is the result of inner transformations (Kogan, 2002) and the view that creative performance involves interaction and tension between the creator's conscious and subconscious which impact the creator's identity (Sawyer, 1992).

## Acknowledgments

We are grateful for funding to the second author from the Natural Sciences and Engineering Research Council of Canada and the Concerted Research Program of the Flemish Government of Belgium. We thank the performers, the Creator's Arts Centre, and Celine Dahmen-Formosa for their assistance, and Gordon Rudolph for comments.